\documentclass[12pt,preprint]{aastex}
\usepackage{amssymb}

\begin{document}

\title{On the Formation of SMC X-1: the Effect of Mass
and Orbital Angular Momentum Loss}

\author{Tao Li and X.-D. Li}
\affil{$^1$Department of Astronomy, Nanjing University, Nanjing 210093,
China}
\affil{$^2$The Key Laboratory of Modern Astronomy and Astrophysics, Ministry of Education,
Nanjing 210093, China}
\email{litao@nju.edu.cn, lixd@nju.edu.cn}

\begin{abstract}
\object{SMC X-1} is a high-mass X-ray binary with an orbital
period of 3.9 days. The mass of the neutron star is as low
as $\sim 1 M_{\sun}$, suggesting that it was likely to be formed through an
electron-capture supernova rather an iron-core collapse supernova.
From the present system configurations, we argue that the orbital
period at the supernova was $\lesssim 10$ days. Since the mass
transfer process between the neutron star's progenitor and the companion
star before the supernova should have increased the orbital period to tens of days, a
mechanism with efficient orbit angular momentum loss and
relatively small mass loss is required to account for its current
orbital period. We have calculated the evolution of the progenitor binary
systems from zero-age main-sequence to the pre-supernova stage
with different initial parameters and various mass and
angular momentum loss mechanisms. Our results show that the outflow
from the outer Langrangian  point or a circumbinary disk formed during the
mass transfer phase may be qualified for this purpose. We point out
that these mechanisms may be popular in binary evolution and significantly
affect the formation of compact star binaries.
\end{abstract}


\section{Introduction}
High-mass X-ray binaries (HMXBs) usually contain an accreting
neutron star (NS) and an early-type (O or B) companion with mass
exceeding $\sim 10 M_\odot$. The NS is fed by the stellar wind or
beginning Roche-lobe overflow (RLOF) from the companion. Since the
NSs in HMXBs have experienced very little accretion because of their
young ages, their masses should be very close to those at birth.
Eclipsing X-ray binary systems where the X-ray source is a pulsar
can be ideal systems for a dynamical determination of the NS masses.
The measured masses of NSs in HMXBs range from
$1.06^{+0.11}_{-0.10}M_\odot$ for \object{SMC X-1} \citep{mee2007}
to $1.86\pm0.16M_\odot$ for \object{Vela X-1}
\citep{bar2001,qua2003}. Recently \citet{raw2011} presented an
improved method for determining the masses of NSs in eclipsing X-ray
pulsars and applied it to six systems. They found the NS masses of
$1.77\pm0.08 M_\odot$ for \object{Vela X-1} and $1.04\pm0.09M_\odot$
for \object{SMC X-1}.  In this paper, we focus  on \object{SMC X-1}
with a NS mass near the minimum mass limit expected for a NS
produced in a supernova (SN) \citep{hae2002,lat2004}.

\object{SMC X-1} was first detected during a rocket flight
\citep{pri1971}.  The discovery of X-ray eclipses with the
\emph{Uhuru} satellite established the binary nature of \object{SMC
X-1}. The pulsar has a pulse period of 0.71 s \citep{luc1976}. The
companion to the X-ray pulsar, \object{Sk 160}, is a B0 I supergiant
located in the ``wing" of the Small Magellanic Cloud (SMC) at a
distance of 60 kpc.  Its mass was estimated to be $17.2\pm0.6
M_\odot$ \citep{rey1993}, $16.6\pm0.4 M_\odot$\citep{val2005}, and
$15.7^{+1.5}_{-1.4}M_\odot$ \citep{mee2007}, while more recent work
by \citet{raw2011} gives $15.35\pm1.53 M_\odot$.  The X-ray source
exhibits an eclipse duration of $0.610\pm0.019$ day \citep{pri1976} in
a 3.892 day orbit. Timing studies of the X-ray pulsations
\citep{lev1993} give $a_{\rm X}\sin i=53.4876\pm0.0004$ ltsec for
the projected semi-major axis and indicate a circular orbit with an
eccentricity $e < 0.00004$. A decay in the orbital period at a rate
$\dot{P}_{\rm orb}/P_{\rm orb} = (-3.36 \pm 0.02) \times 10^{-6}
{\rm yr^{-1}}$ was found \citep{lev1993}, probably due to tidal
interaction between the orbit and the rotation of the companion
star, which is supposed to be in the hydrogen shell burning phase.
The observed period change may also be explained by the
self-sustaining mass loss through the outer Lagrangain ($L_2$) point
\citep{phi2002}. The X-ray emission from \object{SMC X-1} has also
been found to exhibit a long quasi-stable super-orbital period of
$40-60$ days, which is believed to be a result of obscuration of the NS by a
warped, precessing accretion disk \citep{woj1998,cla2003}. In X-rays
both low- and high-intensity states have been observed with an X-ray
luminosity $L_{\rm X}$ varying from $\sim 10^{37} {\rm ergs^{-1}}$
to $\sim 5\times10^{38} {\rm ergs^{-1}}$ \citep{sch1972}. The mass
transfer in \object{SMC X-1} probably has significant contribution
from RLOF \citep{par1984}, as the stellar winds from \object{Sk 160}
are not strong enough to power the X-rays \citep{ham1984}. The X-ray
pulsar in the system has a short spin period (seconds) compared to
those in wind-fed systems (minutes), as the mass and
angular-momentum accretion rate in RLOF systems is much higher than
in the latter. Since its discovery, observations with various
X-ray telescopes  show a steady spin-up of the NS. This makes
\object{SMC X-1} an exceptional X-ray pulsar in which no spin-down
episode has been observed \citep{kah1999}. \citet{li1997} suggested
that the magnetic moment in \object{SMC X-1} may be as low as that of
the bursting pulsar \object{GRO J1744$-$28}, i.e. $\sim 10^{29} {\rm
Gcm^3}$.

Recent analysis by \citet{mee2007} and \citet{raw2011},
like several earlier investigations, found a low value for the NS
mass $\sim 1 M_\odot$. Both theories and statistical analyses
suggest that the initial masses of NSs might have a bimodal underlying
distribution, probably originating from two different formation mechanisms: core
collapse supernovae (CCSNe) and electron-capture supernovae (ECSNe)
\citep[e.g.,][]{pfa2002,heu2004,pod2004,pod2005,sch2010,k10,zha2011,ozel2012}.
A CCSN occurs when a high-mass star develops a degenerate iron
core that exceeds the Chandrasekhar limit, while an
ECSN is associated with the collapse of a lower-mass ONeMg core as
it loses pressure support owing to the sudden capture of electrons
by the Mg and Ne nuclei \citep{nom1984}. Numerical simulations indicate
that ECSNe  are under-energetic compared to
CCSNe by at least an order of magnitude. However they are, at
the same time, relatively fast, neutrino-driven, delayed explosions.
The short time available for the propagation of the shock is
expected to prevent a strong build-up of core asymmetries, leading
to typical kicks of $\lesssim 100\,{\rm km s^{-1}}$, lower than
those in CCSN events \citep{des2006,kit2006}. As
ECSNe should produce somewhat less massive NSs
($\lesssim 1.3 M_\odot$) than CCSNe
\citep{nom1984}, the low mass of the NS in \object{SMC X-1} would be
consistent with its formation by an ECSN in a
degenerate ONeMg core, which is is expected to form at the end of
the evolution of stars with masses of  $\sim 8-13M_\odot$
\citep{pod2005,kit2006}. Thus, assuming an
explosive mass loss during the formation of the NS of about $1
M_\odot$ and a few solar mass of stellar wind mass loss from the NS
progenitor, \citet{mee2007} concluded that a $13 M_\odot +9 M_\odot$
main-sequence progenitor system would be consistent with the
present configuration of \object{SMC X-1}. The more massive star
(i.e., the progenitor of the NS) first evolves to fill its RL,
triggering mass transfer to the less massive star. Since the initial
mass ratio is not extreme, the binary is unlikely to undergo
common envelope (CE) evolution before the SN that cause its orbit to
shrink rapidly \citep[for reviews of CE evolution,
see][]{taa2000,iva2013}. However, a potential problem with such a
scenario is that quasi-conservation of mass and orbital angular momentum
during the mass transfer would lead to a fairly wide pre-SN system,
such that the present orbital period of $\sim 3.9$ days would be
hard to understand, unless a large amount of orbital angular
momentum has been lost with relatively small mass (at most a few
solar masses) from the system. In this paper, we will try to search
plausible evolutionary tracks for \object{SMC X-1}, taking into
account various kinds of orbital angular momentum loss mechanisms.

This paper is organized as follows. In Section 2 we describe the
stellar evolution code and the angular momentum loss mechanisms
used. We present the
calculated results of the binary evolution in Section 3. Discussion and
conclusions are given in Section 4.

\section{Calculations of Binary Evolution}
\subsection{The Stellar Evolution Code}
We adopt an updated version of the stellar evolution code originally
developed by \citet{egg1971,egg1972} to compute the binary
evolution. The binary system is initially composed of two zero-age
main-sequence (ZAMS) stars with an orbital period $P_{\rm orb}$. We set the
more massive star as the primary (of mass $M_1$), which fills the RL
first and is the progenitor of the NS, and the other one as the
secondary (of mass $M_2$). Low metallicities ($Z = 0.004$ and 0.01) are
taken for each star according to the environment of the SMC
\citep{pia2012}. The effective radius $R_{L,1}$ of the RL for the primary star
is calculated from the \citet{egg1983}'s equation,
\begin{equation}
\frac{R_{L,1}}{a} = \frac{0.49q^{2/3}}{0.6q^{2/3}+\ln(1+q^{1/3})},
\end{equation}
where $q = M_1/M_2$ is the mass ratio of the binary components and $a$ is the orbital separation.
The rate of mass transfer via RLOF is calculated with
$\dot{M}_1 = {\rm RMT}\times\max[0, (R_1/R_{L,1}-1)^3]M_\odot {\rm yr^{-1}}$
in the code, where $R_1$ is the radius of the primary, and we adopt ${\rm RMT} = 10^3$
in the calculation.

\subsection{Mass and Angular Momentum Loss Mechanisms}
We assume that the primary star rotates synchronously with the binary orbital revolution,
since the timescale of tidal synchronization is generally much shorter than the characteristic
evolutionary timescale of the binaries considered here. We consider two kinds of mechanisms
of angular momentum loss.
The first is the angular momentum loss due to gravitational radiation. This becomes important when the
orbital period is short. The rate of the angular momentum loss is given by \citep{lan1975}
\begin{equation}
\frac{dJ_{\rm GR}}{dt} = -\frac{32}{5} \frac{G^{7/2}}{c^5}
\frac{M_1^2M_2^2(M_1+M_2)^{-1/2}}{a^{7/2}} ,
\end{equation}
where $J$, $G$, and $c$ are the orbital angular momentum,
gravitational constant, and speed of light, respectively.
The second and more important angular momentum loss mechanism is nonconservative mass transfer.
During the mass-transfer processes, part of the transferred mass from the primary star may
escape the binary system, carrying away the orbital angular momentum \citep{warner78,egg00}.
This mass loss might be related to the rapid rotation of the accreting star.
Since the accreting matter carries a certain angular momentum that will be transferred to the
secondary \citep{p81}, this angular momentum spins up the top layers of
the secondary star, and is transferred further into the star due to rotationally induced mixing processes.
When the secondary is spun up to close to critical rotation it
starts losing mass due to the influence of centrifugal force \citep[e.g.][]{langer98},
although it is not clear in which ways the material leaves the binary.
Considering the complicated processes of
mass loss, we adopt four kinds of
mass loss as follows. In the first case we assume that a fraction $\alpha$
of the transferred mass is  ejected out of the binary
as isotropic winds from the secondary, carrying away the specific angular momentum of the secondary,
\begin{equation}
\frac{dJ_{\rm ml,1}}{dt} = -\alpha \dot{M}_1 \left(\frac{q}{1+q}\right)^2 a^2 \omega ,
\end{equation}
where $\omega$ is the angular velocity of the binary.
Alternatively, the transferred mass might be lost at the inner Lagrangian ($L_1$) point and
the corresponding rate of angular momentum loss is
\begin{equation}
\frac{dJ_{\rm ml,2}}{dt} = -\beta \dot{M}_1 a_{L1}^2 \omega ,
\end{equation}
where $\beta$ is the fraction of mass loss, and
$a_{L1}$ is the distance from the $L_1$ point to the center of mass of the binary system.

In some cases, the matter surrounding the two components may expand to the outer Lagrangian
($L_2$) point, and part of the material lost from the donor star may escape the system through the
$L_2$ point \citep{dre1995,van1998}. Assuming that a fraction $\epsilon$ of the
matter flow leaves the binary,  the angular
momentum loss rate due to the $L_2$ point outflow is given by
\begin{equation}
\frac{dJ_{\rm ml,3}}{dt} = -\epsilon \dot{M}_1 a_{L2}^2 \omega ,
\end{equation}
where $a_{L2}$ is the distance between the mass center of binary and the $L_2$ point.
In our calculation, we take into account the $L_2$ point outflow after the mass ratio inverts,
to ensure that
the outflows is on the side of the lower-mass star that fills the RL.

The last case is the angular momentum loss due to a circumbinary
(CB) disk. CB disks were first proposed by \citet{heu1973} when
investigating the evolution of X-ray binaries. Recently, CB disks
are considered in a wide variety of astrophysical objects,
e.g., young binary stars, protoplanetary systems, cataclysmic
variables (CVs) and massive binary black hole systems in active
galactic nuclei \citep[][and references
therein]{spr2001,hay2009,sl2012}.
It has been argued that during mass exchange in binary
systems, some of the lost matter which possesses high orbital
angular momentum may form a disk surrounding the binary system
rather than leave the binary system \citep{heu1994}.
As shown by \citet{spr2001} and
\citet{che2006}, CB disks can efficiently extract orbit angular
momentum from the binary, and enhance the mass transfer rates. In
this work, we assume that a fraction $\delta$ of the mass lost from the donor
feeds into the CB disk. At the inner edge $r_{\rm i}$ of the disk
tidal torques are then exerted on the binary  via gravitational
interaction, and the corresponding angular momentum loss rate is
\citep{spr2001}
\begin{equation}
\frac{dJ_{\rm CB}}{dt} = -\gamma \left(\frac{2\pi a^2}{P_{\rm orb}}\right)
\dot{M}_{\rm CB} \left(\frac{t}{t_{\rm vi}}\right)^{1/3} ,
\end{equation}
where  $\dot{M}_{\rm CB}=\delta\dot{M}_1$, $\gamma^2 = r_{\rm i}/a = 1.7$
\citep{art1994}, $t$ is the time since the onset of the
CB disk formation, and $t_{\rm vi}$ is the viscous timescale at the inner edge of the CB disk.
For the standard $\alpha$-viscosity disk \citep{ss1973},
\begin{equation}
t_{\rm vi} = \frac{2\gamma^3P_{\rm orb}}{3\pi\alpha_{\rm SS}(H_{\rm i}/r_{\rm i})^2},
\end{equation}
where $H_{\rm i}$ is the scale height of the disk at the inner edge.
In the following  calculation we set the viscosity parameter
$\alpha_{\rm SS} = 0.01$, and assume that the disk is
hydrostatically supported, and geometrically thin with $H_{\rm
i}/r_{\rm i} \sim 0.03$ \citep{bel2004}.


\section{The Formation of SMC X-1}
The formation of a HMXB like SMC X-1 requires two initially massive
stars. When it evolves to become a supergiant and fill its RL, the
primary star transfers material to the secondary on a thermal
timescale, and the binary orbit shrinks rapidly at first.
Because the mass
ratio is not far from unity, the mass transfer process is
dynamically stable without the occurrence of a CE.
A large fraction of the transferred matter
is expected to be accreted by the secondary star. When the mass
ratio inverts, further mass transfer leads to increase of the
orbital period. Finally the primary star develops an ONeMg core,
which collapses to be a NS. At this time the primary star is much
less massive than the secondary. Considering the fact that a small
fraction of the binary mass is lost and a small kick is imparted to
the newborn NS during the SN, the binary orbit
should be nearly circular or mildly eccentric after the SN.
During the subsequent
evolution, since stellar winds of the main-sequence secondary is
not strong, there is hardly any mass transfer until the
secondary starts to fill its RL.
So we expect that the masses of the NS and the optical companion has
changed little since the SN, and will use
them to constrain the binary parameters before the SN. It is noted
that, after the onset of the secondary RLOF, the large mass ratio
between the companion star and the NS will cause the orbit to shrink
in response to mass transfer, most likely leading to a CE event.

\subsection{The effect of the SN kick}
When a star explodes as a SN, the NS formed receives a velocity kick
due to any asymmetry in the explosion \citep{lyn1994}. This also
imparts an impulse to the companion star resulting in an eccentric
binary orbit. As mentioned earlier, the low mass of the NS in
\object{SMC X-1} suggests that its formation is likely to be due to an
electron-capture collapse with a low kick velocity, which results
in a small eccentricity. Tidal interaction between the
NS and the companion star can also circularize the binary
orbit if it is narrow enough.

We first try to estimate the possible distributions of the orbital
period before the SN and the eccentricity after the SN from
current binary parameters. The NS mass ($\sim 1M_\odot$), the
secondary mass ($\sim 16M_\odot$) and the orbital period ($\sim
4$ d) have already been estimated from the observations of
\object{SMC X-1}. The mass of the secondary is assumed to be
constant during the SN, and the primary's mass before the SN can
be constrained  by the range of the helium core mass ($\sim
1.37M_{\odot}-2.5M_{\odot}$) that leads to an ECSN
\citep{nom1984,pod2004}. The SN is thought to take place in a
circular orbit because of the previous mass transfer, and the kick direction
is assumed to be uniform over all directions. We take the kick velocity
$v_{\rm k}$ from a Maxwellian distribution
\begin{equation}
P(v_{\rm k}) = \sqrt{\frac{2}{\pi}}\frac{v_{\rm k}^2}{\sigma_{\rm
k}^3}{\rm e}^{-v_{\rm k}^2/2\sigma_{\rm k}^3},
\end{equation}
where the velocity dispersion $\sigma_{\rm k}$ is usually taken to be $190\,
{\rm kms^{-1}}$ \citep{han1997} or $265\,{\rm
kms^{-1}}$ \citep{hob2005} based on the analysis of various pulsar
proper motion samples for CCSN events. The
kick velocities must be much lower for ECSNe
\citep[e.g.,][]{pfa2002}. A small
kick value ($10\%$ of the standard NS kick, i.e., $\sigma_{\rm k} =
26.5$ kms$^{-1}$) was adopted by \citet{lin2009}, in order to estimate
the overconcentration of ECSN sources in the SMC bar.
Here we set $\sigma_{\rm k} = 20$, 50, and 70 kms$^{-1}$ to examine the
effect of ECSN kicks. Figure 1 shows the derived
distributions of the pre-SN orbital period (top panel) and the post-SN
eccentricity (middle panel) for a  $2M_\odot + 16M_\odot$ pre-SN system, with the
assumption that the post-SN period is $\sim 4$ days. The
results with the CCSN kicks are also presented for
comparison. With the increase of $\sigma_{\rm k}$, a more extended
range of the pre-SN orbital period can result in the 4 day
post-SN period, and the post-SN eccentricity tends to be
larger. For sufficiently large kicks the eccentricity can be larger
than unity so that the binary is disrupted. For CCSNe this occurs at
the high-end in the distribution of the pre-SN
period, providing an upper limit of $\sim 12$ days. This means that a
$> 12$ day pre-SN binary would not evolve to a
4 day post-SN binary with the \object{SMC X-1} mass parameters. From the
contours in the orbital period vs. eccentricity plane (the bottom panel), one can
see the trend of pre-SN period and post-SN eccentricity distributions. There
are two ``wings" with the center close to the post-SN period. The most
probable regions vary from the center to the wings when $\sigma_{\rm
k}$ is increasing. For CCSN kicks, to form the expected post-SN
systems like \object{SMC X-1} requires a  pre-SN period $\lesssim 2$
days, and the post-SN eccentricity is high ($\gtrsim 0.4$). For
ECSNe a lower eccentricity is expected and a
relatively longer pre-SN period ($\sim 2\,{\rm d} - 8\,{\rm d}$) is
required.

\subsection{The Pre-SN Binary Evolution}
Following the above analysis, we simplify the formation of
\object{SMC X-1} to be the evolution of
a ZAMS binary system to a pre-SN binary with the
orbital period $\sim 2\,{\rm d} - 8\,{\rm d}$, the primary (He star)'s mass
$\sim 2 M_\odot$, and the secondary's mass $\sim 15-18 M_\odot$.
We select two
sets of mass parameters for the progenitor binary to illustrate
the possible evolutionary tracks of \object{SMC X-1}. They are $9.98
M_\odot +8.02 M_\odot$ and $12.56 M_\odot +5.60 M_\odot$ ZAMS
binaries with similar initial total masses for the
conservative evolution. We adjust the initial mass of the
secondary if the evolutionary model is
nonconservative, in order to keep the secondary mass at the SN within
the suitable range.

\subsubsection{The binary evolution with $M_{\rm 1,i}=9.98\,M_\odot$}
We first consider the evolution of the progenitor system
with the initial primary's mass $M_{\rm 1,i}=9.98 M_\odot$.
To examine the possible
evolutionary tracks  and the influence
of different angular momentum loss mechanisms, we construct five
models with various kinds of mass and angular momentum loss mentioned in
Section 2.2: (1) conservative mass transfer; (2)
nonconservative mass transfer with mass loss from the secondary; (3)
nonconservative mass transfer with mass loss from the $L_1$ point; (4)
nonconservative mass transfer with mass loss from the $L_2$ point;
and (5)
nonconservative mass transfer with a small fractional mass
feeding into the CB disk. The initial orbit periods are taken to be
$P_{\rm orb, i} = 1.5$, 2.0, 2.5, 3.0, 3.5, 4.0, 5.0, 7.0, and 10.0 days.
At the smallest orbital period the zero-age primary star
just fits the size of the RL.
The metallicities are adopted to be $Z=0.01$
and 0.004.

The results of the pre-SN evolution for different initial
orbital periods are summarized in Tables 1 (for models [1]-[4]),
2, and 3 (for model [5]). In the tables $M_{\rm 1,f}$, $M_{\rm 2,f}$, and $P_{\rm orb, f}$
are the  primary (He star)'s mass, the secondary's mass and the orbital
period close to the end of the primary's evolution, respectively.
In Table 1 we distinguish Cases A and B mass transfer,
which occurs when the primary burns hydrogen in its core and
evolves off the main-sequence but before helium core
ignition, respectively. The Case B mass transfer is most likely to
be unstable if the value of $\epsilon$ is high for model (4).
This is what ``$\dot{M}$ divergent" means in the tables. So we only present
the results of model (2)-(4) with Case A and stable Case B
mass transfer.

The results of the first three models in Table 1 show little
distinction.
The orbital angular momentum loss from the binaries is all
inefficient, so that the final orbital periods are usually
larger than 40 days (for $Z=0.01$) or 25 days (for $Z=0.004$),
inconsistent with those for
\object{SMC X-1}. It is also noted that a larger $Z$ leads to longer
$P_{\rm orb,f}$  because of longer mass transfer time. For each model, the
higher $P_{\rm orb,i}$, the heavier the evolved He star
and the lighter the secondary. It is interesting to note that
$P_{\rm orb,f}$ decreases with increasing $P_{\rm orb,i}$
for Case A mass transfer, but increases with $P_{\rm orb,i}$ for Case B mass transfer.
This means the lowest pre-SN period $P_{\rm orb,f}$ corresponds
to the $P_{\rm orb,i}$ around the boundary between
Case A and Case B evolution.
Different from the first three models,
in model (4) the mass loss from the
$L_2$ point can cause secular orbital shrinking. If the mass loss
fraction $\epsilon\sim 25\%$, $P_{\rm orb,f}$ becomes
shorter than 10 days, which satisfies what is needed for the
formation of \object{SMC X-1}.

Although the $L_2$ point outflow provides a way for the formation of relatively
compact systems. this mechanism requires a considerable mass loss
rate, and sometimes the primary's mass at the end of the
evolution is lower than the range for an ECSN.
Thus in model (5) we take into account the influence of
a CB disk formed during the mass transfer. In this case
we adopt the progenitor binary system of $9.98
M_\odot +8.12 M_\odot$ with the secondary's mass slightly higher than
that in model (1), and we only calculate the evolution with $P_{\rm orb,i}$
near the the boundary between Case A and Case B mass transfer. Similar as
in model (4), the Case B mass transfer becomes unstable if
$\delta$ is large, so we mainly focus on Case A evolution.
The results are presented in Tables 2 and 3 with $Z=0.01$ and
$Z=0.004$, respectively. We can always find that it is able to
form a pre-SN binary system with a relatively small
$\delta$ (a few hundredth),  which meets the \object{SMC
X-1} requirements. The wider the initial system, the
higher $\delta$ needed.

We then select some representative scenarios from the tables
to demonstrate the detailed evolution of the binary. Figure~2
shows the pre-SN evolutionary tracks of the primary of initial mass
$9.98 M_\odot$ in the H-R diagram. The
panels from top to bottom correspond to models (1)-(5),
respectively. In the left, middle and right columns
the initial orbital period $P_{\rm orb,i}$ is taken to be 2.0,
2.5, and 3.0 days,
respectively. For model (1), the initial secondary's mass is
$M_{\rm 2,i}=8.02 M_\odot$;
for models (2)-(4), we adopt $\alpha=\beta=\epsilon = 0.25$;
for model (5), $M_{\rm 2,i} =
9.62 M_\odot$, and $M_{\rm 2,i} =8.12 M_\odot$.
The solid
and dotted lines are obtained with $Z = 0.01$ and
0.004, respectively. In Fig.~3, we compare the mass transfer rates
between different models with different $P_{\rm orb, i}$. The figure is
organized same as Fig.~2. The only Case B evolution is for
$P_{\rm orb,i}= 3.0$ days and $Z =0.004$, and in other scenarios are
all Case A evolutions. We can see that there are not any
significant differences in the evolutionary tracks and in the mass
transfer rates among the first three models with the same
$P_{\rm orb, i}$, while the other two models are distinct from the
others. The binaries generally experience three (or two,
the first two phases are sometimes degenerated) phases of mass
transfer in Case A evolution in Fig.~3. In the first phase
the mass transfer proceeds on a thermal timescale \citep[see][for
further discussions on thermally unstable mass transfer]{lan2000}.
The outcome in this situation is that the donor star overfills its
RL even more, leading to further mass loss. Although the
star is driven out of thermal equilibrium during this phase, it
manages to retain hydrostatic equilibrium and the system can in this
case avoid a so-called delayed dynamical instability
\citep{hje1987,kal1996}, which would have resulted in a CE stage.
The second and third phases proceed on
nuclear timescales determined by the core burning of the remaining
hydrogen and, later on, the hydrogen shell burning, respectively.
Obviously the wider the
initial system, the later the RLOF mass transfer begins, because
there is no effective angular momentum loss mechanism before the
onset of mass transfer, and the mass transfer timescale becomes
shorter. When  $P_{\rm orb, i}$ is so long that there is Case B
mass transfer,
the three phases merge into one. Furthermore,
lower metallicity causes later onset of RLOF and
shorter mass transfer time due to a larger initial stellar
radius and ending the red giant stage earlier. For models (4) and
(5), if $\epsilon$ or $\delta$ is high enough, the mass transfer can be dynamically
unstable, so that the binary enters the CE evolution
when the orbital period is  as low as $\sim 2$ days.

The orbital period
evolution is shown in Fig.~4, which is also organized similar as
Fig.~2, but in each subgraph we add the cases with different
values of the mass loss fraction in different colored lines for comparison. The dashed
horizontal lines indicate the possible range of the pre-SN periods
expected to lead to the formation of \object{SMC X-1}.
It is clearly seen that only in model (4) or (5) with suitable values
of $\epsilon$ or $\delta$ can the binary evolve into the proper orbital period.

\subsubsection{The binary evolution with $M_{\rm 1,i}=12.56\,M_\odot$}
In this subsection we investigate the evolution of the possible
progenitor of \object{SMC X-1} with a $12.56\,M_\odot$ main-sequence
primary. Tables 4, 5 and 6 present the results of the
binary parameters for the primary evolving
from ZAMS to pre-SN in different models, similar as those for the
$9.98 M_\odot$ primary. The orbital period evolution is shown
in Fig.~5. Compared with the case of $M_{\rm 1,i}=9.98 M_\odot$,
the higher mass ratio for $M_{\rm 1,i}=12.56 M_\odot$ leads to shorter
$P_{\rm orb,f}$ for similar total mass, initial orbital period and metallicity.
Usually models (4) and (5) are more preferred, while in the case of
conservative mass transfer, the $P_{\rm orb,f}$
sometimes nearly fits in the pre-SN period range
for the formation of \object{SMC X-1}, but
the He star mass exceeds the
upper limit of the He core mass for an ECSN.

Combining the calculated results of binary evolutions with different
initial parameters, we can roughly constrain the mass of the NS progenitor in
\object{SMC X-1} to be $\sim 10 - 12.5 M_\odot$.

\section{Discussion and Conclusions}
Since the pioneering work of \citet{hh72} the formation and evolution of
HMXBs with NS components has been intensively studied
\citep{rh82,mh89,p91,l94,ds95,ity95,tts98,lzw11}.
The generally accepted evolutionary picture is that,
the originally more massive star becomes the less
massive as a result of mass transfer or mass loss, leaving behind a system
composed of an evolved core and a MS companion.
The subsequent evolution of the core to the
SN leads to the formation of a NS/massive star binary system.
However, there are big uncertainties associated
with the effects of mass and angular momentum loss on the
evolution of these systems (in the context of the mass transfer
and CE evolution). Detailed investigations on the formation of
individual HMXBs are also lacking because it is difficult to constrain
the initial binary parameters from current observations.
The low mass of the NS in \object{SMC X-1} is suggestive of
its formation through an ECSN, which provides useful
information on the properties of the NS progenitor and the
formation scenarios. ECSNe are thought to occur
in stars of mass $\sim 8-13\,M_{\sun}$. Combining with the total mass
of the current binary, this suggests that  the initial masses of the
primary and secondary stars are comparable so that CE evolution
is not likely to occur, and the orbital evolution is mainly
dominated by the mass transfer/loss processes and tidal interaction.
If the mass transfer is conservative before
the SN, the ratio between the final and initial orbital periods is
\begin{equation}
\frac{P_{\rm orb,f}}{P_{\rm orb,i}}=\left(\frac{M_{\rm 1,i}M_{2,i}}
{M_{\rm 1,f}M_{2,f}}\right)^3\sim 50-100.
\end{equation}
This gives an orbital period at least tens of days when the SN occurs
(the orbit may shrink to some extent due to the wind from the primary),
and in the further evolution it would be difficult for the orbital period to
decrease to $\sim 4$ days by tidal torques within the lifetime of the secondary star.
This point has already been notified by \citet{rl98} to explain
why Be/X-ray binaries are in wide ($>10$ days) and eccentric orbits.
For \object{SMC X-1}, as the orbital change during the post-SN
evolution is limited, through numerical calculation of the change
in the orbital period during the SN, we simplify the evolutional problem
to evolving a
progenitor ZAMS binary system up to the pre-SN epoch, when it has developed
an ONeMg core approximately fitting the range expected
to lead to an ECSN, with the orbital period  $\lesssim 8-10$ days,
and the secondary mass $\sim 15-17\,M_{\sun}$.
Obviously these requirements rule out the conservative mass transfer
scenario, and the key problem is to find an effective orbital angular loss
mechanism with relatively small mass loss to cause secular orbital shrinking.
According to our calculation, it could be either mass loss
through the $L_2$ point or or a CB disk (note that the formation of a
CB disk might also originate from the outflow from the $L_2$ point). The
difference is that, in the former case some of the material leaves the system
carrying away the orbital angular momentum, while in the latter
the lost mass accumulates in a disk surrounding the binary,
draining the orbital angular momentum from the binary through
tidal torques. Because of the difference in the efficiency of angular momentum
loss, the fraction of lost mass in the total transferred mass is
significantly higher for the $L_2$ point outflow than for the CB disk.
Our calculated results indicate that the evolution with the $L_2$
point mass loss or a CB disk could lead to the formation of \object{SMC X-1}
through Case A mass transfer ($P_{\rm orb,i} \lesssim 3.5$ days) with
$M_{\rm 1,i}\sim 10 - 12.5 M_\odot$.

Although our analysis focuses on the formation of \object{SMC X-1}, the
conclusion that mass transfer is likely to be associated with significant
angular momentum loss may be a general feature in the evolution of
various kinds of binaries \citep[e.g.,][]{r74,my75,s93,egg00,p05,p08}.
Previous investigations emphasized the necessity of nonconservative
mass transfer to account for the properties of the related binaries.
However, the ways of mass loss and the corresponding efficiency of angular
momentum loss have not been well constrained. In this work we show that the
outflow from the $L_2$ point or a circumbinary disk is more preferred
than mass loss from the accreting star or the $L_1$ point, at least for the
formation of \object{SMC X-1}.

Similar suggestions can be found in the literature. For example,
in order to address the observed lower
mass limit ($\sim 8\,M_{\sun}$) for the Be stars in Be/X-ray binaries,
\citet{pz95} proposed that there may be mass loss from the binary systems
at the $L_2$ point when the primary star transfers mass to the secondary
(who's rotation will be accelerated such that it will become a Be star).
The escaped matter takes away about six times
the specific angular momentum of the binary system, so that
those with small initial mass ratios would undergo spiral-in and evolve towards
a CE phase.
In the other extreme of low-mass binaries,
\citet{kni2011} recently reconstructed the complete evolutionary path followed
by cataclysmic variables (CVs), using the observed mass-radius relationship of the
secondary stars. The best-fit revised model of CV evolution
indicates that the angular momentum loss rate below the period gap is
$2.47(\pm 0.22)$ times the rate induced by gravitational radiation,
suggesting the existence of some other angular momentum loss mechanisms.
\citet{sl2012} considered several kinds of consequential angular momentum loss
mechanisms  including
isotropic wind from the accreting white dwarfs, outflows from the Langrangian
points, and the formation of a CB disk. They showed that neither isotropic
wind from the white dwarf nor outflow from the $L_1$ point can explain the extra
angular momentum loss rate, while outflow from the $L_2$ point or a
CB disk can satisfy the extra angular momentum loss provided that
$\sim (15-45)\%$ of the transferred mass is lost from the binary, or $\lesssim 10^{-3}$
of the transferred mass goes into the CB disk \citep[see also][]{spr2001}.
These results are in general line with what we have obtained for \object{SMC X-1}.

Mass transfer critically affects the evolutionary paths of
binaries. Depending on whether it is stable or not, a
binary may either survive the initial mass transfer phase to
become a semi-detached system or  end up merging completely.
The stability of mass transfer depends sensitively on the
angular momentum transport mechanisms. Including the mass loss
from the $L_2$ point (or a CB disk) can significantly destabilize
the mass transfer processes, enhance the merging rate,
and influence the birth rate of compact star binaries.
Obviously a thorough investigation on this subject is needed before
it can be implemented into future population synthesis calculation.

This work was supported by the Natural Science Foundation of China under grant numbers
11133001 and 11333004, the National Basic Research Program of China
(973 Program 2009CB824800),
and the Qinglan project of Jiangsu Province.

\clearpage

\begin{figure}
\plotone{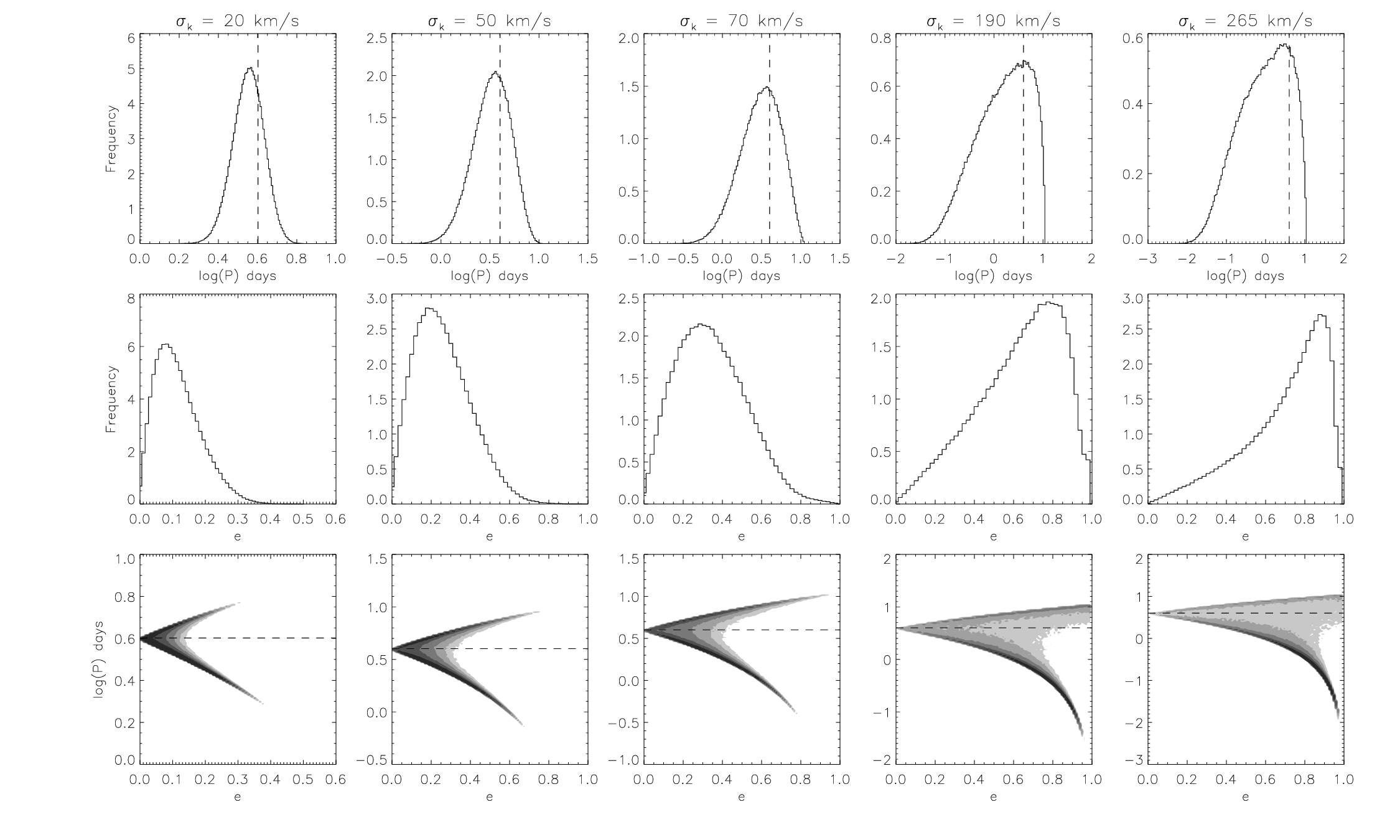} \caption{Distributions of the pre-SN orbital period
(top) and the post-SN eccentricity  (middle) for a binary consisting
of a $2M_\odot$ He star and a $16M_\odot$ MS star with $\sigma_{\rm k} = 20,
50, 70, 190, 265\,{\rm km s^{-1}}$. The mass of the newborn NS is
assumed to be $1M_\odot$. In the bottom panels are the contours
representing the relative probabilities of the system parameters in
the pre-SN orbital period vs. the post-SN eccentricity plane. The darker the color,
the greater the probabilities. The expected post-SN period ($\sim 4$
days) is also drawn with the dashed line. \label{fig1}}
\end{figure}

\begin{figure}
\epsscale{.80}
\plotone{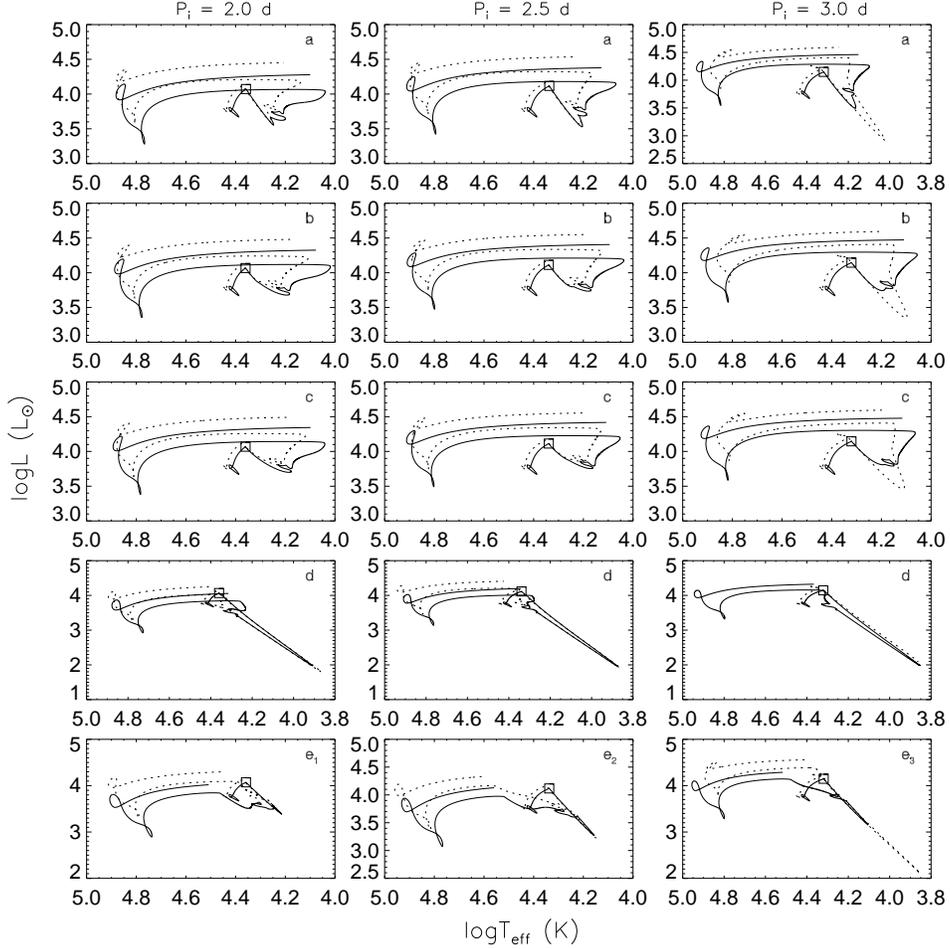} \caption{The evolutionary tracks for the
primary star of mass $9.98 M_\odot$ from ZAMS to the formation of
the O-Ne-Mg star in the H-R diagram. The initial orbital periods
are 2.0 days (left), 2.5 days (middle), and 3.0 days (right).
The top to bottom panels correspond to models (1) to (5), and the
solid and dotted lines represent the results with
$Z = 0.01$ and $Z = 0.004$, respectively. The open squares
correspond to the onset of RLOF. The model parameters are as follows.
a: $M_{\rm
2,i}=8.02M_\odot$; b: $M_{\rm 2,i}=9.62M_\odot$,
$\alpha=0.25$; c: $M_{\rm 2,i}=9.62M_\odot$, $\beta=0.25$;
d: $M_{\rm 2,i}=9.62M_\odot$, $\epsilon=0.25$; e$_1$:
$M_{\rm 2,i}=8.12M_\odot$, $\delta=0.015$; e$_2$: $M_{\rm
2,i}=8.12M_\odot$, $\delta=0.025$ for $Z = 0.01$ and $\delta=0.03$
for $Z = 0.004$; e$_3$: $M_{\rm 2,i}=8.12M_\odot$,
$\delta=0.03$ for $Z = 0.01$ and $\delta=0.08$ for $Z = 0.004$.
\label{fig2}}
\end{figure}

\begin{figure}
\plotone{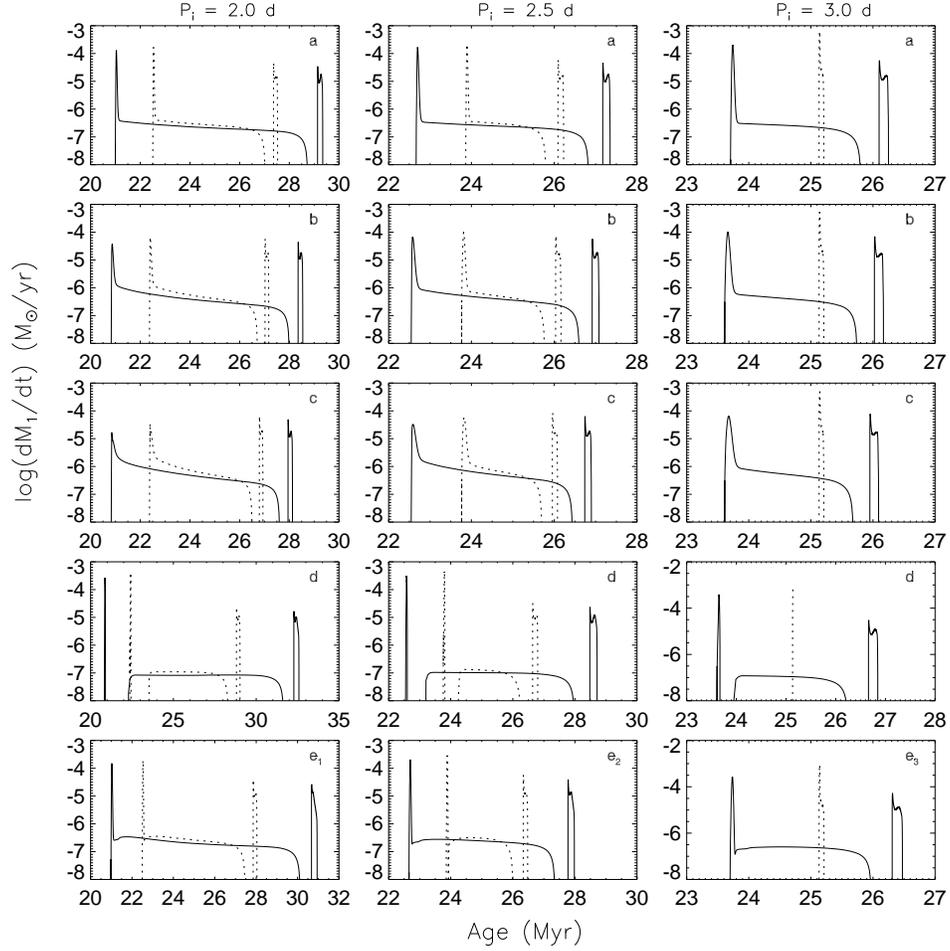}
\caption{Evolution of the mass transfer rate $\dot{M}_1$ for different
binary models. The initial orbital periods are $P_{\rm i}=$ 2 days (left),
3 days (middle), and 4 days (right). Other symbols have the same meanings as
in Fig.~2. \label{fig3}}
\end{figure}

\begin{figure}
\epsscale{.70} \plotone{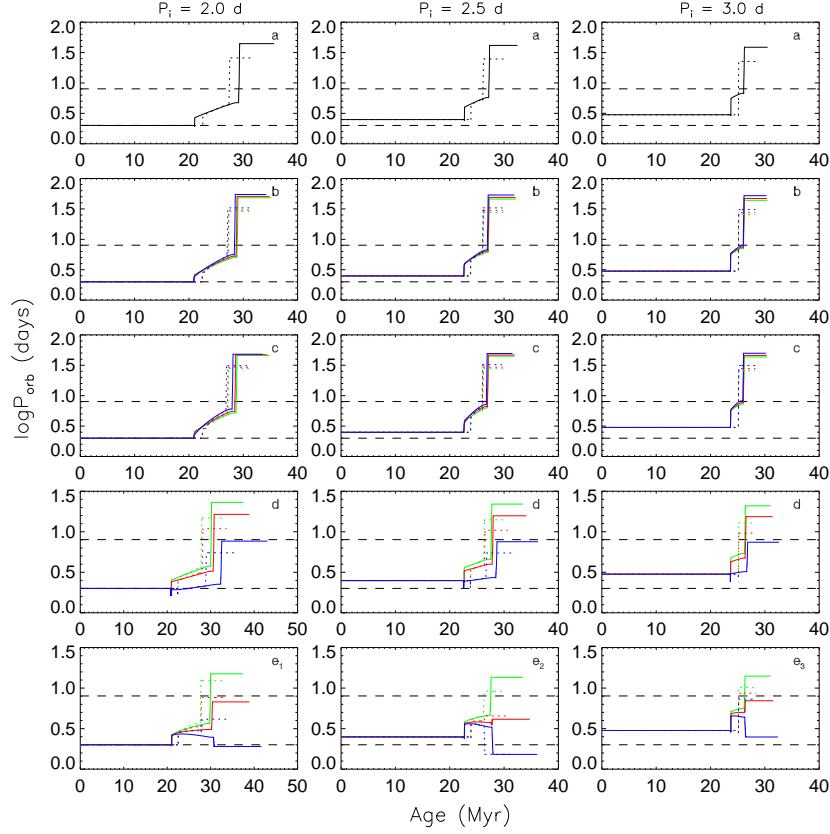} \caption{Evolution of the orbital
period for a ZAMS binary with the primary
mass of $9.98 M_\odot$.
The top to bottom panels correspond to models (1) to (5).
The initial orbital periods are $P_{\rm i}=$ 2 days (left),
3 days (middle), and 4 days (right), and the solid and
dotted lines represent $Z = 0.01$ and $Z = 0.004$, respectively.
Other model parameters are as follows. a:
$M_{\rm 2,i}=8.02M_\odot$; b: $M_{\rm 2,i}=8.62M_\odot$,
$\alpha=0.1$ ({\it green}), $M_{\rm 2,i}=9.02M_\odot$, $\alpha=0.15$
({\it red}), and $M_{\rm 2,i}=9.62M_\odot$, $\alpha=0.25$ ({\it
blue}); c: $M_{\rm 2,i}=8.62M_\odot$, $\beta=0.1$ ({\it
green}), $M_{\rm 2,i}=9.02M_\odot$, $\beta=0.15$ ({\it red}), and
$M_{\rm 2,i}=9.62M_\odot$, $\beta=0.25$ ({\it blue}); d:
$M_{\rm 2,i}=8.62M_\odot$, $\epsilon=0.1$ ({\it green}), $M_{\rm
2,i}=9.02M_\odot$, $\epsilon=0.15$ ({\it red}), and $M_{\rm
2,i}=9.62M_\odot$, $\epsilon=0.25$ ({\it blue}); $\rm e_1$:
$M_{\rm 2,i}=8.12M_\odot$, $\delta=0.01$ ({\it green}),
$0.015$ ({\it red}) and $0.02$ ({\it blue}); $\rm e_2$: $M_{\rm
2,i}=8.12M_\odot$, $\delta=0.015$ ({\it green}), $0.025$
({\it red}), $0.03$ ({\it blue}) for $Z = 0.01$ and
$\delta=0.02$ ({\it green}), $0.03$ ({\it red}),
$0.04$ ({\it blue}) for $Z = 0.004$; $\rm e_3$:
$M_{\rm 2,i}=8.12M_\odot$, $\delta=0.04$ ({\it green}),
$0.05$ ({\it red}), $0.06$ ({\it blue}) for $Z = 0.01$
and $\delta=0.06$ ({\it green}), $0.07$ ({\it red}),
$0.08$ ({\it blue}) for $Z = 0.004$. The dashed horizontal
lines indicate the approximate range of the pre-SN period that can be
expected to lead to the formation of SMC X-1. \label{fig4}}
\end{figure}

\begin{figure}
\epsscale{.80}
\plotone{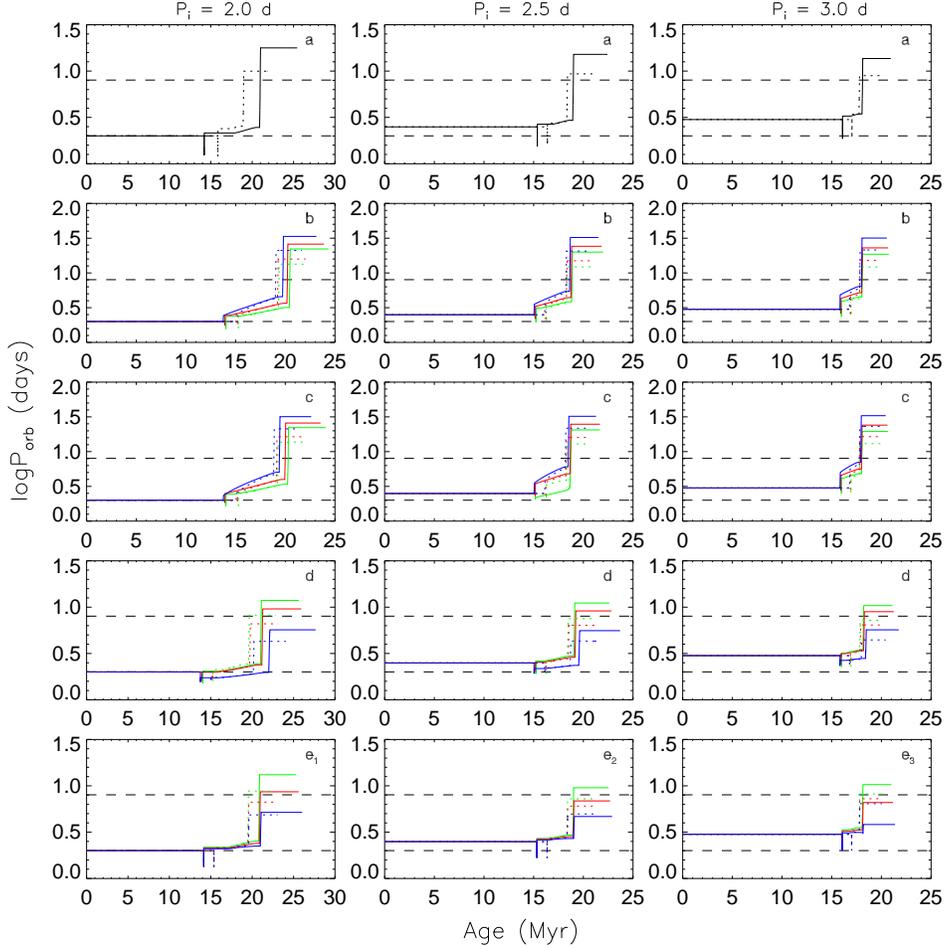}
\caption{Same as Fig.~4, but for a ZAMS binary with the primary
mass of $12.56 M_\odot$. The model parameters are as follows.
a: $M_{\rm 2,i}=5.60M_\odot$;
b-d: $M_{\rm 2,i}=7.00M_\odot$, $\alpha=\beta=\epsilon=0.1$ ({\it green}),
$M_{\rm 2,i}=8.00M_\odot$, $\alpha=\beta=\epsilon=0.15$ ({\it red}),
and $M_{\rm 2,i}=10.00M_\odot$,
$\alpha=\beta=\epsilon=0.25$ ({\it blue});
$\rm e_1$: $M_{\rm 2,i}=6.00M_\odot$, $\delta=0.005$ ({\it green}),
$0.01$ ({\it red}) and $0.015$ ({\it blue});
$\rm e_2$: $M_{\rm 2,i}=6.00M_\odot$, $\delta=0.01$ ({\it green}), $0.015$ ({\it red}) and $0.02$
({\it blue});
$\rm e_3$: $M_{\rm 2,i}=6.00M_\odot$, $\delta=0.01$ ({\it green}),
$0.02$ ({\it red}), $0.03$ ({\it blue}) for $Z = 0.01$ and $\delta=0.015$
({\it green}), $0.025$ ({\it red}), $0.035$ ({\it blue}) for $Z = 0.004$.
\label{fig5}}
\end{figure}

\begin{deluxetable}{cccccccccccc}
\tabletypesize{\scriptsize} \rotate \tablecaption{Calculated Results
of the Binary Evolution with $M_{\rm 1,i}=9.98 M_\odot$ in models
(1)-(4). \label{tbl-1}} \tablewidth{0pt}
\tablehead{
 & & & & & & $Z=0.01$ & & & & $Z=0.004$ & \\
$P_{\rm orb,i}$(days) & $M_{\rm 2,i}$($M_\odot$) & mass loss fraction & Model &
Case  & $M_{\rm 1,f}$($M_\odot$)
& $M_{\rm 2,f}$($M_\odot$) & $P_{\rm orb,f}$(days) &  Case  & $M_{\rm 1,f}$($M_\odot$) & $M_{\rm 2,f}$($M_\odot$) & $P_{\rm orb,f}$(days)\\
 & & $\alpha$, $\beta$, $\gamma$, or $\epsilon$ & & & & & & & & & \\
} \startdata
1.5 & 8.02 & 0    & 1 & A & 1.53 & 16.47 & 48.38 & A & 1.85 & 16.15 & 28.70 \\
    & 8.62 & 0.10 & 2 & A & 1.57 & 16.19 & 51.42 & A & 1.90 & 15.89 & 30.82 \\
    &      &      & 3 & A & 1.59 & 16.17 & 47.63 & A & 1.92 & 15.87 & 29.40 \\
    &      &      & 4 & A & 1.38 & 16.43 & 24.93 & A & 1.69 & 16.16 & 15.94 \\
    & 9.02 & 0.15 & 2 & A & 1.60 & 16.14 & 53.29 & A & 1.93 & 15.86 & 32.11 \\
    &      &      & 3 & A & 1.63 & 16.12 & 47.55 & A & 1.96 & 15.84 & 29.86 \\
    &      &      & 4 & A & 1.29 & 16.48 & 17.68 & A & 1.59 & 16.22 & 11.64 \\
    & 9.62 & 0.25 & 2 & A & 1.65 & 15.87 & 56.23 & A & 1.98 & 15.62 & 34.06 \\
    &      &      & 3 & A & 1.69 & 15.83 & 47.02 & A & 2.03 & 15.58 & 30.33 \\
    &      &      & 4 & A & 1.07 & 16.39 &  8.50 & A & 1.36 & 16.16 &  5.82 \\
2.0 & 8.02 & 0    & 1 & A & 1.75 & 16.25 & 44.30 & A & 2.15 & 15.85 & 25.82 \\
    & 8.62 & 0.10 & 2 & A & 1.79 & 15.99 & 48.25 & A & 2.19 & 15.64 & 28.48 \\
    &      &      & 3 & A & 1.81 & 15.98 & 45.95 & A & 2.20 & 15.63 & 27.98 \\
    &      &      & 4 & A & 1.63 & 16.21 & 23.06 & A & 2.00 & 15.87 & 14.74 \\
    & 9.02 & 0.15 & 2 & A & 1.81 & 15.96 & 50.81 & A & 2.21 & 15.63 & 30.17 \\
    &      &      & 3 & A & 1.84 & 15.94 & 47.06 & A & 2.22 & 15.61 & 29.29 \\
    &      &      & 4 & A & 1.55 & 16.26 & 16.38 & A & 1.91 & 15.95 & 10.92 \\
    & 9.62 & 0.25 & 2 & A & 1.85 & 15.72 & 54.55 & A & 2.24 & 15.43 & 32.93 \\
    &      &      & 3 & A & 1.89 & 15.69 & 48.04 & A & 2.24 & 15.43 & 32.19 \\
    &      &      & 4 & A & 1.36 & 16.14 &  7.66 & A & 1.69 & 15.88 &  5.51 \\
2.5 & 8.02 & 0    & 1 & A & 1.96 & 16.04 & 41.43 & A & 2.39 & 15.61 & 24.77 \\
    & 8.62 & 0.10 & 2 & A & 1.98 & 15.82 & 45.94 & A & 2.41 & 15.43 & 27.71 \\
    &      &      & 3 & A & 1.99 & 15.81 & 44.63 & A & 2.41 & 15.43 & 27.67 \\
    &      &      & 4 & A & 1.84 & 16.01 & 21.94 & A & 2.28 & 15.62 & 14.02 \\
    & 9.02 & 0.15 & 2 & A & 2.00 & 15.80 & 48.84 & A & 2.42 & 15.44 & 29.64 \\
    &      &      & 3 & A & 2.02 & 15.79 & 46.61 & A & 2.43 & 15.44 & 29.47 \\
    &      &      & 4 & A & 1.78 & 16.07 & 15.72 & A & 2.18 & 15.73 & 10.71 \\
    & 9.62 & 0.25 & 2 & A & 2.03 & 15.58 & 53.53 & A & 2.44 & 15.27 & 32.86 \\
    &      &      & 3 & A & 2.06 & 15.56 & 49.25 & A & 2.46 & 15.26 & 32.34 \\
    &      &      & 4 & A & 1.59 & 15.96 &  7.53 & A & 2.01 & 15.73 &  5.47 \\
3.0 & 8.02 & 0    & 1 & A & 2.15 & 15.85 & 38.63 & B & 2.63 & 15.37 & 23.27 \\
    & 8.62 & 0.10 & 2 & A & 2.16 & 15.65 & 43.83 & B & 2.67 & 15.20 & 25.54 \\
    &      &      & 3 & A & 2.17 & 15.65 & 43.33 & B & 2.67 & 15.20 & 25.89 \\
    &      &      & 4 & A & 2.05 & 15.82 & 20.90 & B & 2.61 & 15.32 & 12.83 \\
    & 9.02 & 0.15 & 2 & A & 2.18 & 15.65 & 46.92 & B & 2.68 & 15.23 & 27.50 \\
    &      &      & 3 & A & 2.18 & 15.65 & 45.92 & B & 2.68 & 15.22 & 27.94 \\
    &      &      & 4 & A & 1.99 & 15.90 & 15.34 & B & 2.53 & 15.43 &  9.84 \\
    & 9.62 & 0.25 & 2 & A & 2.19 & 15.46 & 52.20 & B & 2.69 & 15.09 & 30.82 \\
    &      &      & 3 & A & 2.21 & 15.45 & 49.82 & B & 2.69 & 15.09 & 31.39 \\
    &      &      & 4 & A & 1.83 & 15.79 &  7.42 & B & &$\dot{M}$ divergent&\\
3.5 & 8.02 & 0    & 1 & A & 2.36 & 15.64 & 35.83 & B & 2.68 & 15.32 & 26.00 \\
    & 8.62 & 0.10 & 2 & A & 2.36 & 15.48 & 41.01 & B & 2.69 & 15.18 & 29.44 \\
    &      &      & 3 & A & 2.36 & 15.48 & 40.94 & B & 2.69 & 15.18 & 29.86 \\
    &      &      & 4 & A & 2.28 & 15.63 & 19.70 & B & 2.63 & 15.30 & 14.73 \\
    & 9.02 & 0.15 & 2 & A & 2.36 & 15.50 & 44.28 & B & 2.69 & 15.22 & 31.71 \\
    &      &      & 3 & A & 2.37 & 15.49 & 43.99 & B & 2.69 & 15.21 & 32.24 \\
    &      &      & 4 & A & 2.23 & 15.68 & 14.33 & B & 2.59 & 15.38 & 11.01 \\
    & 9.62 & 0.25 & 2 & A & 2.36 & 15.33 & 49.63 & B & 2.70 & 15.08 & 35.55 \\
    &      &      & 3 & A & 2.37 & 15.33 & 49.11 & B & 2.70 & 15.08 & 36.24 \\
    &      &      & 4 & A & 2.10 & 15.51 &  7.19 & B & &$\dot{M}$ divergent&\\
4.0 & 8.02 & 0    & 1 & B & 2.52 & 15.48 & 34.49 & B & 2.69 & 15.31 & 29.41 \\
    & 8.62 & 0.10 & 2 & B & 2.53 & 15.33 & 39.16 & B & 2.70 & 15.17 & 33.32 \\
    &      &      & 3 & B & 2.53 & 15.33 & 39.51 & B & 2.70 & 15.17 & 33.80 \\
    &      &      & 4 & B & 2.47 & 15.44 & 18.99 & B & 2.64 & 15.29 & 16.67 \\
    & 9.02 & 0.15 & 2 & B & 2.53 & 15.35 & 42.26 & B & 2.70 & 15.21 & 35.89 \\
    &      &      & 3 & B & 2.53 & 15.35 & 42.63 & B & 2.70 & 15.20 & 36.50 \\
    &      &      & 4 & B & 2.44 & 15.50 & 13.97 & B & 2.61 & 15.36 & 12.47 \\
    & 9.62 & 0.25 & 2 & B & 2.54 & 15.20 & 47.52 & B & 2.71 & 15.07 & 40.24 \\
    &      &      & 3 & B & 2.54 & 15.20 & 47.81 & B & 2.71 & 15.07 & 41.06 \\
    &      &      & 4 & B & &$\dot{M}$ divergent&& B & &$\dot{M}$ divergent&\\
5.0 & 8.02 & 0    & 1 & B & 2.54 & 15.46 & 42.47 & B & 2.71 & 15.29 & 36.16 \\
7.0 & 8.02 & 0    & 1 & B & 2.56 & 15.44 & 58.20 & B & 2.73 & 15.27 & 49.43 \\
10.0& 8.02 & 0    & 1 & B & 2.58 & 15.42 & 81.51 & B & 2.76 & 15.24 & 69.03 \\
\enddata
\end{deluxetable}

\clearpage

\begin{deluxetable}{cccccc}
\tabletypesize{\scriptsize} \tablecaption{Calculated Results
of the Binary Evolution in model (5) with $M_{\rm 1,i}=9.98 M_\odot$,
$M_{\rm 2, i}=8.12M_\odot$ and $Z=0.01$. \label{tbl-2}}
\tablewidth{0pt} \tablehead{
$P_{\rm orb,i}$(days) & Model & mass loss fraction $\delta$ & $M_{\rm 1,f}$($M_\odot$)
& $M_{\rm 2,f}$($M_\odot$) & $P_{\rm orb,f}$(days) \\
} \startdata
2.0 & 5 & 0.01 & 1.61 & 16.42 & 15.02 \\
    &   & 0.015& 1.51 & 16.48 &  6.76 \\
    &   & 0.02 & 1.34 & 16.60 &  1.91 \\
2.5 & 5 & 0.015& 1.81 & 16.18 & 13.58 \\
    &   & 0.02 & 1.75 & 16.21 &  8.06 \\
    &   & 0.025& 1.66 & 16.26 &  4.13 \\
    &   & 0.03 & 1.53 & 16.34 &  1.52 \\
3.0 & 5 & 0.02 & 2.03 & 15.93 & 14.03 \\
    &   & 0.025& 1.99 & 15.93 & 10.15 \\
    &   & 0.03 & 1.94 & 15.95 &  6.98 \\
    &   & 0.035& 1.87 & 15.98 &  4.48 \\
    &   & 0.04 & 1.79 & 16.02 &  2.49 \\
3.5 & 5 & 0.03 & 2.27 & 15.63 & 13.55 \\
    &   & 0.04 & 2.23 & 15.60 & 10.06 \\
    &   & 0.05 & 2.18 & 15.58 &  6.70 \\
    &   & 0.06 & 2.12 & 15.57 &  4.20 \\
    &   & 0.07 & 2.04 & 15.58 &  2.42 \\
\enddata
\end{deluxetable}

\clearpage

\begin{deluxetable}{cccccc}
\tabletypesize{\scriptsize} \tablecaption{Calculated Results
of the Binary Evolution in model (5) with $M_{\rm 1,i}=9.98 M_\odot$,
$M_{\rm 2, i}=8.12M_\odot$ and $Z=0.004$. \label{tbl-3}}
\tablewidth{0pt} \tablehead{
$P_{\rm orb,i}$(days) & Model & mass loss fraction $\delta$ & $M_{\rm 1,f}$($M_\odot$) &
$M_{\rm 2,f}$($M_\odot$) & $P_{\rm orb,f}$(days) \\
} \startdata
2.0 & 5 & 0.005& 2.09 & 15.98 & 18.44 \\
    &   & 0.01 & 2.01 & 16.02 & 12.36 \\
    &   & 0.015& 1.92 & 16.08 &  7.66 \\
    &   & 0.02 & 1.80 & 16.15 &  4.13 \\
    &   & 0.025& 1.64 & 16.27 &  1.88 \\
2.5 & 5 & 0.01 & 2.32 & 15.72 & 15.85 \\
    &   & 0.02 & 2.23 & 15.74 &  9.17 \\
    &   & 0.03 & 2.10 & 15.79 &  4.54 \\
    &   & 0.04 & 1.89 & 15.92 &  1.52 \\
3.0 & 5 & 0.04 & 2.62 & 15.22 & 13.57 \\
    &   & 0.05 & 2.60 & 15.18 & 11.77 \\
    &   & 0.06 & 2.58 & 15.14 & 10.14 \\
    &   & 0.07 & 2.55 & 15.09 &  8.64 \\
    &   & 0.08 & 2.54 & 15.04 &  7.31 \\
    &   & 0.09 & &$\dot{M}$ divergent&\\
\enddata
\end{deluxetable}

\clearpage

\begin{deluxetable}{cccccccccccc}
\tabletypesize{\scriptsize} \rotate \tablecaption{Calculated Results
of the Binary Evolution with $M_{\rm 1,i}=12.56 M_\odot$ in models
(1)-(4). \label{tbl-4}} \tablewidth{0pt}
\tablehead{
 & & & & & & $Z=0.01$ & & & & $Z=0.004$ & \\
$P_{\rm orb,i}$(days) & $M_{\rm 2,i}$($M_\odot$) & mass loss fraction & Model &
Case  & $M_{\rm 1,f}$($M_\odot$)
& $M_{\rm 2,f}$($M_\odot$) & $P_{\rm orb,f}$(days) &  Case  & $M_{\rm 1,f}$($M_\odot$) & $M_{\rm 2,f}$($M_\odot$) & $P_{\rm orb,f}$(days)\\
 & & $\alpha$, $\beta$, $\gamma$, or $\epsilon$ & & & & & & & & & \\
} \startdata
2.0 & 5.60 & 0    & 1 & A & 2.11 & 16.05 & 17.79 & A & 2.70 & 15.46 &  9.98 \\
    & 7.00 & 0.10 & 2 & A & 2.25 & 16.28 & 23.14 & A & 2.79 & 15.79 & 13.38 \\
    &      &      & 3 & A & 2.33 & 16.21 & 22.14 & A & 2.83 & 15.75 & 13.72 \\
    &      &      & 4 & A & 2.09 & 16.71 & 11.82 & A & 2.56 & 16.26 &  7.50 \\
    & 8.00 & 0.15 & 2 & A & 2.37 & 16.66 & 25.97 & A & 2.88 & 16.23 & 15.85 \\
    &      &      & 3 & A & 2.42 & 16.62 & 25.81 & A & 2.92 & 16.19 & 16.37 \\
    &      &      & 4 & A & 2.05 & 17.27 &  9.57 & A & 2.51 & 16.88 &  6.61 \\
    & 10.0 & 0.25 & 2 & A & 2.51 & 17.54 & 33.54 & A & 3.00 & 17.17 & 21.06 \\
    &      &      & 3 & A & 2.59 & 17.48 & 31.86 & A & 3.07 & 17.11 & 21.26 \\
    &      &      & 4 & A & 1.87 & 18.36 &  5.69 & A & 2.30 & 18.10 &  4.20 \\
    & 11.50& 0.50 & 2 & A & 2.64 & 16.46 & 39.75 & A & 3.13 & 16.21 & 25.43 \\
    &      &      & 3 & A & 2.80 & 16.38 & 36.58 & A & 3.29 & 16.14 & 25.72 \\
    &      &      & 4 & A & &$\dot{M}$ divergent&& A & &$\dot{M}$ divergent&\\
2.5 & 5.60 & 0    & 1 & A & 2.46 & 15.70 & 15.14 & A & 2.93 & 15.23 &  9.34 \\
    & 7.00 & 0.10 & 2 & A & 2.60 & 15.96 & 19.94 & A & 3.17 & 15.45 & 12.24 \\
    &      &      & 3 & A & 2.63 & 15.93 & 20.52 & A & 3.20 & 15.43 & 12.89 \\
    &      &      & 4 & A & 2.41 & 16.41 & 10.90 & A & 2.95 & 15.92 &  7.20 \\
    & 8.00 & 0.15 & 2 & A & 2.66 & 16.42 & 24.20 & A & 3.22 & 15.94 & 14.96 \\
    &      &      & 3 & A & 2.70 & 16.38 & 24.82 & A & 3.25 & 15.92 & 16.01 \\
    &      &      & 4 & A & 2.38 & 17.00 &  9.11 & A & 2.88 & 16.57 &  6.51 \\
    & 10.00& 0.25 & 2 & A & 2.77 & 17.34 & 32.39 & A & 3.31 & 16.93 & 20.61 \\
    &      &      & 3 & A & 2.83 & 17.30 & 32.27 & A & 3.35 & 16.90 & 21.73 \\
    &      &      & 4 & A & 2.20 & 18.90 &  5.58 & A & 2.72 & 17.73 &  4.31 \\
    & 11.50& 0.50 & 2 & A & 2.87 & 16.34 & 39.90 & A & 3.39 & 16.09 & 25.94 \\
    &      &      & 3 & A & 2.99 & 16.28 & 39.39 & A & 3.48 & 16.04 & 28.42 \\
    &      &      & 4 & A & &$\dot{M}$ divergent&& A & &$\dot{M}$ divergent&\\
3.0 & 5.60 & 0    & 1 & A & 2.75 & 15.41 & 13.67 & A & 3.23 & 14.93 &  8.93 \\
    & 7.00 & 0.10 & 2 & A & 2.88 & 15.71 & 18.60 & A & 3.43 & 15.22 & 12.19 \\
    &      &      & 3 & A & 2.90 & 15.69 & 19.50 & A & 3.44 & 15.20 & 13.02 \\
    &      &      & 4 & A & 2.71 & 16.15 & 10.39 & A & 3.28 & 15.63 &  7.21 \\
    & 8.00 & 0.15 & 2 & A & 2.92 & 16.20 & 22.93 & A & 3.45 & 15.74 & 15.19 \\
    &      &      & 3 & A & 2.95 & 16.17 & 24.07 & A & 3.47 & 15.72 & 16.43 \\
    &      &      & 4 & A & 2.67 & 16.75 &  8.93 & A & 3.24 & 16.26 &  6.43 \\
    & 10.00& 0.25 & 2 & A & 3.00 & 17.17 & 31.77 & A & 3.51 & 16.79 & 21.35 \\
    &      &      & 3 & A & 3.03 & 17.15 & 32.91 & A & 3.54 & 16.77 & 23.09 \\
    &      &      & 4 & A & 2.50 & 17.88 &  5.69 & A & 3.07 & 17.46 &  4.43 \\
    & 11.50& 0.50 & 2 & A & 3.05 & 16.25 & 40.53 & A & 3.55 & 16.00 & 27.55 \\
    &      &      & 3 & A & 3.16 & 16.20 & 41.99 & A & 3.62 & 15.97 & 31.20 \\
    &      &      & 4 & A & &$\dot{M}$ divergent&& A & &$\dot{M}$ divergent&\\
3.5 & 5.60 & 0    & 1 & A & 3.01 & 15.15 & 12.74 & B & 3.48 & 14.68 &  8.81 \\
    & 7.00 & 0.10 & 2 & A & 3.11 & 15.51 & 17.96 & B & 3.64 & 15.03 & 12.35 \\
    &      &      & 3 & A & 3.13 & 15.49 & 18.99 & B & 3.66 & 15.01 & 13.25 \\
    &      &      & 4 & A & 2.98 & 15.90 & 10.04 & B & &$\dot{M}$ divergent&\\
    & 8.00 & 0.15 & 2 & A & 3.14 & 16.00 & 22.27 & B & 3.68 & 15.54 & 15.19 \\
    &      &      & 3 & A & 3.17 & 15.98 & 23.78 & B & 3.70 & 15.53 & 16.65 \\
    &      &      & 4 & A & 2.94 & 16.52 &  8.81 & B & &$\dot{M}$ divergent&\\
    & 10.00& 0.25 & 2 & A & 3.18 & 17.03 & 31.75 & B & 3.72 & 16.63 & 21.56 \\
    &      &      & 3 & A & 3.21 & 17.01 & 33.56 & B & 3.74 & 16.62 & 23.81 \\
    &      &      & 4 & A & 2.78 & 17.66 &  5.78 & B & &$\dot{M}$ divergent&\\
    & 11.50& 0.25 & 2 & A & 3.24 & 16.16 & 40.82 & B & 3.75 & 15.90 & 28.11 \\
    &      &      & 3 & A & 3.30 & 16.13 & 44.71 & B & 3.77 & 15.89 & 33.43 \\
    &      & 0.25 & 4 & A & &$\dot{M}$ divergent&& B & &$\dot{M}$ divergent&\\
4.0 & 5.60 & 0    & 1 & B & 3.41 & 14.75 & 10.91 & B & 3.51 & 14.65 &  9.94 \\
    & 7.00 & 0.10 & 2 & B & 3.46 & 15.19 & 15.87 & B & 3.66 & 15.01 & 13.94 \\
    &      &      & 3 & B & 3.48 & 15.17 & 16.99 & B & 3.68 & 14.99 & 14.94 \\
    &      &      & 4 & B & &$\dot{M}$ divergent&& B & &$\dot{M}$ divergent&\\
    & 8.00 & 0.15 & 2 & B & 3.51 & 15.70 & 19.52 & B & 3.70 & 15.53 & 17.19 \\
    &      &      & 3 & B & 3.52 & 15.69 & 21.30 & B & 3.71 & 15.52 & 18.88 \\
    &      &      & 4 & B & &$\dot{M}$ divergent&& B & &$\dot{M}$ divergent&\\
    & 10.00& 0.25 & 2 & A & 3.39 & 16.88 & 31.03 & B & 3.74 & 16.62 & 24.46 \\
    &      &      & 3 & A & 3.41 & 16.86 & 33.43 & B & 3.75 & 16.61 & 27.03 \\
    &      &      & 4 & A & 3.11 & 17.42 &  5.78 & B & &$\dot{M}$ divergent&\\
    & 11.50& 0.50 & 2 & A & 3.41 & 16.07 & 40.85 & B & 3.76 & 15.90 & 31.91 \\
    &      &      & 3 & A & 3.44 & 19.50 & 46.49 & B & 3.78 & 15.89 & 37.99 \\
    &      &      & 4 & A & &$\dot{M}$ divergent&& B & &$\dot{M}$ divergent&\\
5.0 & 5.60 & 0    & 1 & B & 3.44 & 14.72 & 13.35 & B & 3.54 & 14.62 & 12.17 \\
7.0 & 5.60 & 0    & 1 & B & 3.49 & 14.67 & 18.15 & B & 3.59 & 14.57 & 16.55 \\
10.0& 5.60 & 0    & 1 & B & 3.53 & 14.63 & 25.24 & B & 3.63 & 14.53 & 23.01 \\
\enddata
\end{deluxetable}

\clearpage

\begin{deluxetable}{cccccc}
\tabletypesize{\scriptsize} \tablecaption{Calculated Results
of the Binary Evolution in model (5) with $M_{\rm 1,i}=12.56 M_\odot$,
$M_{\rm 2, i}=6.00M_\odot$, and $Z=0.01$. \label{tbl-5}}
\tablewidth{0pt} \tablehead{
$P_{\rm orb,i}$(days) & Model & mass loss fraction $\delta$ & $M_{\rm 1,f}$($M_\odot$) & $M_{\rm 2,f}$($M_\odot$) & $P_{\rm orb,f}$(days) \\
} \startdata
2.0 & 5 & 0.005& 2.12 & 16.41 & 13.18 \\
    &   & 0.01 & 2.06 & 16.43 &  8.63 \\
    &   & 0.015& 1.98 & 16.47 &  5.17 \\
2.5 & 5 & 0.005& 2.46 & 16.07 & 12.74 \\
    &   & 0.01 & 2.41 & 16.08 &  9.54 \\
    &   & 0.015& 2.35 & 16.10 &  6.88 \\
    &   & 0.02 & 2.29 & 16.13 &  4.67 \\
3.0 & 5 & 0.01 & 2.72 & 15.78 & 10.31 \\
    &   & 0.015& 2.68 & 15.78 &  8.32 \\
    &   & 0.02 & 2.63 & 15.79 &  6.61 \\
    &   & 0.025& 2.59 & 15.81 &  5.11 \\
    &   & 0.03 & 2.53 & 15.83 &  3.82 \\
3.5 & 5 & 0.01 & 3.01 & 15.49 & 11.12 \\
    &   & 0.015& 2.99 & 15.48 &  9.84 \\
    &   & 0.02 & 2.97 & 15.47 &  8.67 \\
    &   & 0.025& 2.94 & 15.46 &  7.63 \\
    &   & 0.03 & 2.91 & 15.46 &  6.61 \\
    &   & 0.035& 2.88 & 15.45 &  5.76 \\
    &   & 0.04 & 2.86 & 15.45 &  5.02 \\
\enddata
\end{deluxetable}

\clearpage

\begin{deluxetable}{cccccc}
\tabletypesize{\scriptsize} \tablecaption{Calculated Results
of the Binary Evolution in model (5) with $M_{\rm 1,i}=12.56 M_\odot$,
$M_{\rm 2, i}=6.00M_\odot$, and $Z=0.004$. \label{tbl-3}}
\tablewidth{0pt} \tablehead{
$P_{\rm orb,i}$(days) & Model & mass loss fraction $\delta$ & $M_{\rm 1,f}$($M_\odot$) & $M_{\rm 2,f}$($M_\odot$) & $P_{\rm orb,f}$(days) \\
} \startdata
2.0 & 5 & 0.005& 2.60 & 15.93 &  8.80 \\
    &   & 0.01 & 2.54 & 15.96 &  6.67 \\
    &   & 0.015& 2.46 & 16.00 &  4.86 \\
2.5 & 5 & 0.01 & 2.97 & 15.53 &  7.23 \\
    &   & 0.015& 2.92 & 15.54 &  6.05 \\
    &   & 0.02 & 2.87 & 15.57 &  4.97 \\
    &   & 0.025& 2.81 & 15.59 &  4.02 \\
3.0 & 5 & 0.015& 3.30 & 15.14 &  8.18 \\
    &   & 0.02 & 3.30 & 15.14 &  7.71 \\
    &   & 0.025& 3.28 & 15.13 &  7.27 \\
    &   & 0.03 & 3.27 & 15.11 &  6.84 \\
    &   & 0.035& 3.26 & 15.09 &  6.43 \\
\enddata
\end{deluxetable}




\end{document}